\documentclass[reprint,amsmath,amssymb,aps,prr]{revtex4-1}

\usepackage{graphicx}
\usepackage{dcolumn}
\usepackage{hyperref}
\usepackage[mathlines]{lineno}
\usepackage{soul} 

\usepackage{color}
\begin{document}
\title{Efficient generation of turbulent collisionless shocks in laser-ablated counter-streaming plasmas}

\author{A. Grassi}
\email{anna.grassi@polytechnique.edu}
\affiliation{High Energy Density Science Division, SLAC National Accelerator Laboratory, Menlo Park, California 94025, USA}%
\author{F. Fiuza}
\email{fiuza@slac.stanford.edu}
\affiliation{High Energy Density Science Division, SLAC National Accelerator Laboratory, Menlo Park, California 94025, USA}%

\begin{abstract}
\noindent Laser-ablated high-energy-density (HED) plasmas offer a promising route to study astrophysically relevant processes underlying collisionless shock formation, magnetic field amplification, and particle acceleration in the laboratory. Using large-scale, multi-dimensional particle-in-cell simulations, we explore the interpenetration of laser-ablated counter-streaming plasmas for realistic experimental flow profiles. We find that the shock formation and its structure are substantially different from those of more idealized and commonly considered uniform flows: 
shock formation can be up to 10 times faster due to the transition from small-angle scattering to magnetic reflection and the shock front develops strong corrugations at the ion gyroradius scale. These findings have important consequences for current experimental programs and open exciting prospects for studying the microphysics of turbulent collisionless shocks with currently available high-energy laser systems. 
\end{abstract}

\pacs{Valid PACS appear here}
\maketitle

\section{Introduction}
Collisionless shocks are ubiquitous across a wide range of scales and astrophysical environments, from galaxy clusters~\citep{vanWeeren2017} and supernova remnants ~\citep{Volk2005} to the Earth's bow shock~\citep{Schwartz2011,Johlander2016}. 
They are mediated by plasma instabilities, which dissipate the flow energy by heating the plasma, amplifying magnetic fields, and accelerating particles. The microphysics governing the nonlinear shock dynamics is complex; it cannot be directly resolved by astrophysical observations and is still not fully understood. 
In particular, how the structure of the shock front and the amplification of turbulent magnetic fields affect particle injection into diffusive shock acceleration remains a very important open question in high-Mach number ($M \gg 10$) astrophysical shocks~\cite{Treumann2009, Spitkovsky2008, Caprioli2014, Matsumoto2015, Matsumoto2017, Marle2018,Bohdan2019,Bohdan2020}.

The advent of high-power lasers has opened a new avenue to produce and study high-Mach number collisionless shocks in the laboratory. The main goal is to produce shocks with self-generated magnetic turbulence, where the mechanisms of magnetic field amplification and particle injection can be directly studied in a controlled environment. In particular, there is a strong interest in studies of 
the Weibel instability~\cite{Weibel1959,Fried1959}, as both simulations~\cite{Kato2008,Spitkovsky2008,Matsumoto2017,Lemoine2019,Bohdan2019,Bohdan2020} and spacecraft observations~\cite{Sundberg2017} suggest 
this instability is dominant in driving magnetic turbulence at shocks in a wide range of 
scenarios, from planetary shocks to young supernova remnants to gamma-ray bursts. Experimental studies can greatly complement spacecraft observations and play a critical role in benchmarking numerical simulations and theoretical models. 

This prospect has led to a 
large experimental effort in high-energy-density (HED) laser facilities in the last decade. In these experiments, high-Mach number counter-streaming flows are 
produced from the ablation of a solid-density target by kilojoule, nanosecond laser pulses. Important progress has been achieved in the characterization of the plasma flow properties and interaction~\citep{Takabe2008,Kugland2012,Ross2012,Li2013}, leading to the demonstration of the magnetic field amplification by the Weibel instability~\citep{Fox2013,Huntington2015,Huntington2017,Swadling2020}, and to a better understanding of the interplay between collisional and collisionless effects~\citep{Ross2013,Ross2017}. 

The generation of Weibel-mediated shocks requires a large flow interpenetration distance that poses a significant experimental challenge. Theoretical models and kinetic simulations have been largely limited to uniform plasma flows and suggest that shock formation requires system sizes exceeding $10^3$ ion skin depths~\citep{Kato2008,Ruyer2016,Ruyer2018}, which are beyond those attained at current experimental facilities. However, very recently, experiments at the National Ignition Facility (NIF) have demonstrated the formation of high-Mach number shocks mediated by electromagnetic instabilities~\cite{Fiuza2020}. The shock formation time observed is significantly faster than previous models predict. It is thus crucial to understand the physics of shock formation in the laboratory experiments and in which conditions can current systems allow the study of the relevant microphysics of astrophysical turbulent shocks.

In this paper, we study the formation of collisionless shocks in laser-ablated counter-streaming plasmas using large-scale two- (2D) and three-dimensional (3D) particle-in-cell (PIC) simulations that include the intrinsic density and velocity inhomogeneity of experimental plasma flows. 
We find that realistic plasma profiles impact the interaction in two important ways, which have not been previously recognized: 
(i) shock formation is significantly faster than previously found for homogeneous plasmas because the coherence length of the amplified magnetic field becomes comparable to the ion gyroradius, leading to a transition from small-angle scattering to magnetic reflection that efficiently slows down the flows, and
(ii) enhanced magnetic field advection towards the shock leads to ion-gyroradius-scale corrugations of the shock front. 
An analytical model for the shock formation time and minimum system size required to study collisionless shocks with laser-ablated plasma flows is introduced and shown to be in good agreement with both PIC simulations and recent NIF experiments. 
These results indicate that state-of-the-art HED facilities, such as the NIF and Laser Megajoule (LMJ), can produce collisionless shocks with strong turbulence driven by the Weibel instability and where the shock front uniformity can be controlled, opening the possibility to probe their influence on particle injection and 
test relevant astrophysical shock models.

\section{Plasma flow profiles and simulation setup }
Plasma flows produced by laser-ablation of solid-density targets have inhomogeneous density and velocity profiles given by a well-established self-similar solution~\citep{Gurevich1966}, consistent with experimental measurements, such as those in Refs.~\citep{Ross2012,Ross2013}. In a typical configuration, two counter-facing solid targets separated by a distance $2L_0$ are irradiated by kJ-class lasers to produce counter-streaming plasma flows that interact at the mid-plane between the two targets, defined at $x = 0$. The velocity and density of one of the flows is $v(x,t)= c_{\rm s} + (x+L_0)/(t+\tau_0)$ and $n(x,t)= \tilde{n}\,{\rm exp} \left[- (x+L_0)/(c_{\rm s}(t+\tau_0)) -1\right]$, with $\tau_0$ the time the flow takes to travel the distance $L_0$ to the mid-plane where the two flows first meet at $t=0$, $c_{\rm s}= (\gamma Zk_{\rm B}T_e/m_i)^{1/2}$ the sound speed, $\gamma$ the adiabatic index, $m_i$ and $Z$ the ion mass and charge number, $T_e$ the electron temperature, and $\tilde{n}$ the target surface density. The opposite, counter-propagating flow has the symmetric profile with respect to $x = 0$.

\begin{figure}
    \includegraphics[width=.48\textwidth]{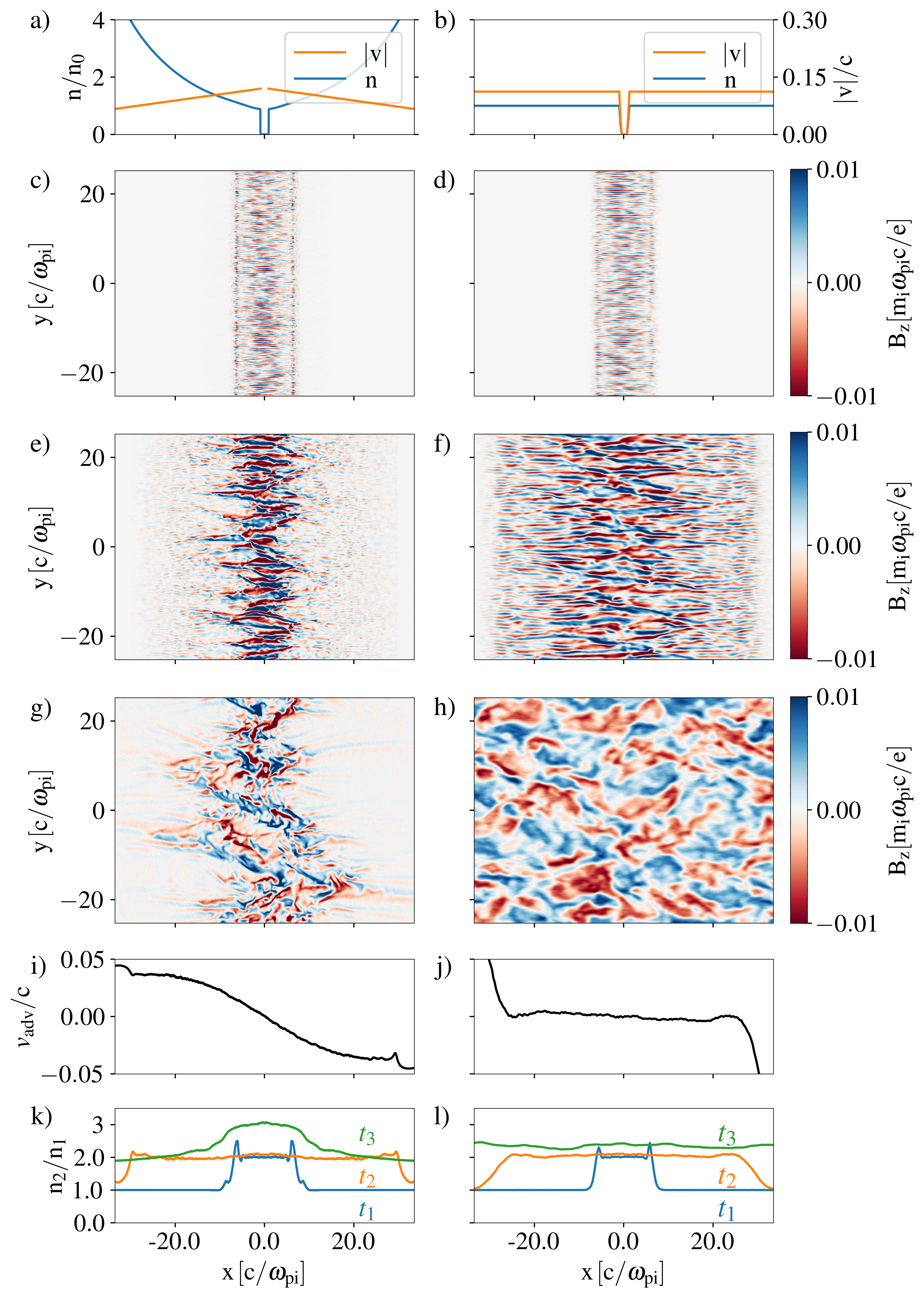}
    \caption{Evolution of the interaction of counter-streaming flows from 2D PIC simulations using laser-ablated inhomogeneous (left column) and homogeneous (right column) profiles. a,b) Initial density (blue) and velocity (orange) profiles. Magnetic field $B_z(x,y)$ evolution at c,d) $t_1\simeq 70\,\omega_{\rm pi}^{-1}$, e,f) $t_2\simeq 285\,\omega_{\rm pi}^{-1}$, and g,h) $t_3\simeq 1780\,\omega_{\rm pi}^{-1}$. i,j) Magnetic field advection velocity ($v_{\rm adv}$) profile at $t_2 \simeq 285\,\omega_{\rm pi}^{-1}$. k,l) Density compression ratio $n_2/n_1$ between the transversely averaged ion density of 
    interacting flows $n_2$ and 
    single flow $n_1$.}
    \label{Figure1}
\end{figure}

In order to study the importance of more realistic plasma profiles on shock formation we have performed 2D and 3D fully kinetic PIC simulations with the code OSIRIS 3.0~\citep{Fonseca2002,Fonseca2008} for both \emph{homogeneous} and \emph{inhomogeneous} counter-streaming non-relativistic plasma flows.
In the homogeneous case (identified by the index H), plasma flows are initialized with uniform velocity $v_{\rm H} = v_0 = 0.11\,c$, a sonic Mach number $M = v_0/c_s = 21$, and density $n_{\rm H} = n_0$. In the inhomogeneous case (identified by the index $\rm I$), 
the flow velocity profile follows the self-similar theory, with maximum velocity $v_0 = 0.11\,c$ near the interaction region (mid-plane of the simulation) at $t = 0$ (Fig.~\ref{Figure1}a). The density profile also closely resembles the self-similar solution (see Fig.~\ref{Figure1}a and Appendix~\ref{app:Init}), but has been modified such that $n_{\rm I}(x=0, t) v_{\rm I}(x=0, t)^2 = n_{\rm H} v_{\rm H}^2 = n_0 v_0^2$. This guarantees that the same plasma energy density is delivered to the interaction region in both cases and allows for a direct comparison of the shock formation efficiency between both. 

The baseline simulations have a longitudinal size of $L_x = 135 \, c/\omega_{\rm pi}$ and a transverse size of $L_y = 50\,c/\omega_{\rm pi}$, with $\omega_{\rm pi} = [4 \pi (2n_0) Z^2 e^2/m_i]^{1/2}$ the ion plasma frequency corresponding to the total plasma density. 
The simulations resolve the electron skin depth at the interaction region ($c/\omega_{\rm pe}$) with 8 cells, use a time step of 0.087 $\omega_{\rm pe}^{-1}$, and use 36 particles per cell per species. The boundary conditions are periodic along the transverse $y$-direction and open along the $x$-axis for both fields and particles. A third-order interpolation scheme is used for improved numerical accuracy. The baseline ion mass to charge ratio used is $m_i/(m_e Z)=128$. The high flow velocity ($\sim0.1\,c$) and reduced mass ratio used are typical choices in the PIC modeling of shocks \cite{Kato2008,Fox2013,Ruyer2015,Ruyer2018} that capture the non-relativistic nature of the flows and a large separation between electron and ion physics, while balancing computational expense. We note that the physics of pure electromagnetic instabilities, such as the Weibel instability discussed in this work, can be rigorously scaled between non-relativistic systems with different flow velocities \citep{Ryutov2012}.

The evolution of the magnetic field and plasma density for both the homogeneous and inhomogeneous cases is shown in Fig.~\ref{Figure1}. 
As expected, the early time dynamics is dominated by the development of the ion Weibel instability and is similar between both cases, which is consistent with previous experiments and 
simulations~\citep{Fox2013,Huntington2015}. The instability gives rise to filamentary currents with transverse wavelength comparable to the ion skin depth $c/\omega_{\rm pi}$ and to the exponential amplification of the magnetic field 
with growth rate 
$\Gamma \simeq 0.07\, \omega_{\rm pi}$ for both cases, in good agreement with linear theory~\citep{Fried1959,Weibel1959}. The instability saturates at $\tau_{\rm sat}\simeq 10\, \Gamma^{-1} \simeq 140\,\omega_{\rm pi}^{-1}$, with $B_{\rm sat} \simeq 0.014 \, m_i\omega_{\rm pi}c/e$, which is 
in good agreement with 
predictions 
based on magnetic trapping 
$B_{\rm sat}=v_0 m_i\omega_{\rm pi}/(2\pi Ze) $ ~\citep{Davidson1972}.

After saturation of the Weibel instability, the transverse coherence length of the magnetic field increases in both cases due to filament merging~\citep{Medvedev2005,AchterbergNL2007,Ruyer2015,Zhou2019}. However, 
the longitudinal extent of the filaments starts to be significantly more limited for the inhomogeneous flows, with the stronger magnetic fields being confined to the central region near the mid-plane 
(Fig.~\ref{Figure1}e,f).
This is a consequence of different effects associated with the inhomogeneity of the flow profiles. The growth rate of the Weibel instability is reduced away from the mid-plane due to the density and velocity asymmetry and due to the ion heating of the flow that has crossed the interaction region (detailed calculations in Appendix~\ref{app:WeibelGrowth}). Moreover, the asymmetric flow profiles lead to enhanced magnetic field advection towards the mid-plane region (Fig.~\ref{Figure1}i,j). The advection velocity is $v_{\rm adv}=(n_{+}v_{+}+n_{-}v_{-})/(n_{+}+n_{-})$~\citep{Ryutov2013}, where the index $+ (-)$ refers to the flow moving in the positive (negative) $x$-direction. (Note that $v_{\rm adv} = 0$ for homogeneous symmetric flows.) The weaker magnetic fields produced in the upstream region at a distance $L$ from the mid-plane are thus advected to the central region and compressed on a time scale $\tau_{\rm adv}= L/v_{\rm adv}$.

\begin{figure}
    \centering
    \includegraphics[width=0.5\textwidth]{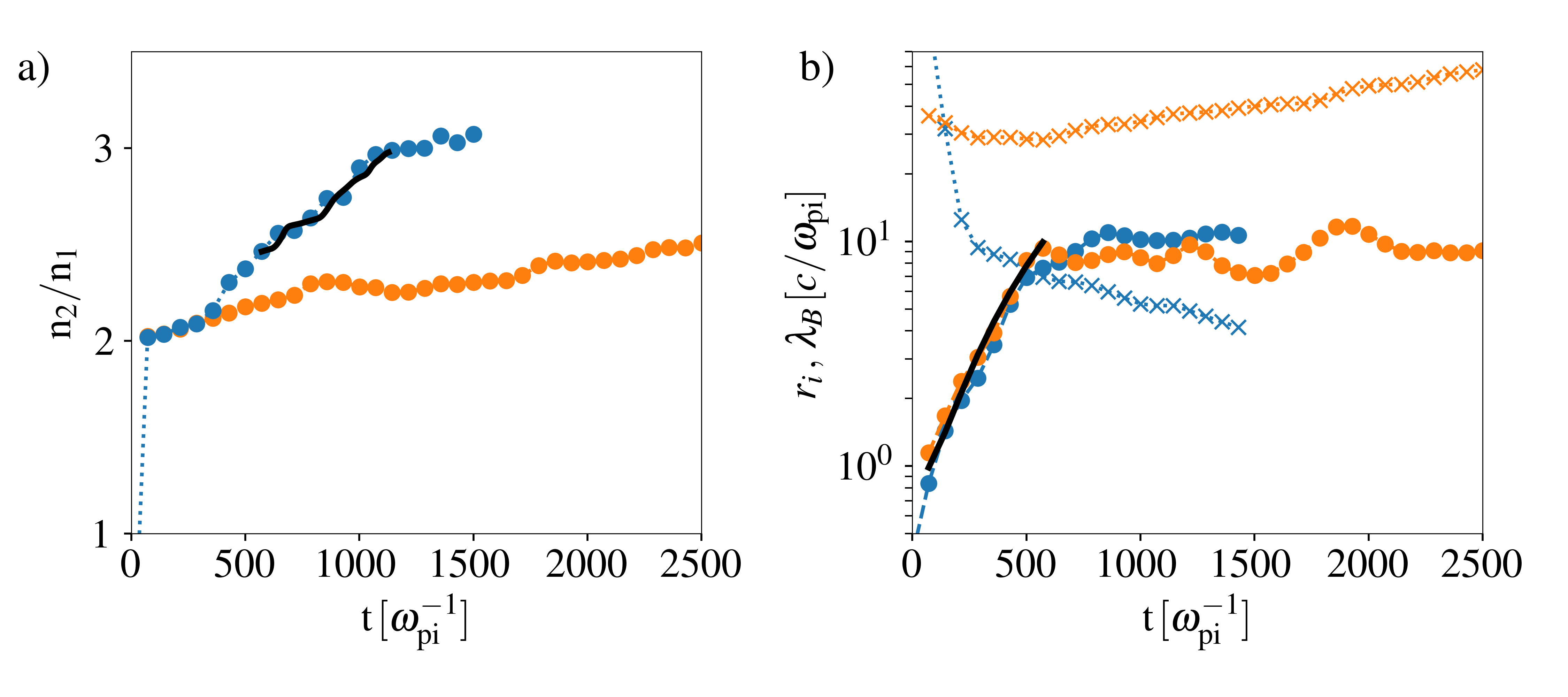}
    \caption{a) Temporal evolution of the density compression ratio $n_2/n_1$ at the simulation mid-plane ($x=0$) for the inhomogeneous (blue dots) and homogeneous (orange dots) flow profiles. The black curve corresponds to the density evolution predicted from Eq.~\ref{equation1} for 
    $t > \tau_c$. b) Temporal evolution of the ion gyroradius $r_{i}$ (crosses) and dominant magnetic wavelength $\lambda_{B}$ (dots) for the inhomogenous (blue) and homogeneous (orange) flow profiles. The black curve corresponds to the filament merging model for $\lambda_{B}(t)$ from Ref.~\citep{Ruyer2015}.}
    \label{Figure2}
\end{figure}

At later times, we observe dramatic differences between the two cases, both in terms of fields (Fig.~\ref{Figure1}g,h) and density (Fig.~\ref{Figure1}k,l). In the case of laser-ablated inhomogeneous plasmas, the flows are strongly compressed in the interaction region, two collisionless shocks are formed  and propagate against the incoming plasma flows. Each shock reaches the density compression dictated by the Rankine-Hugoniot jump conditions~\citep{Zeldovich} $n_2/n_1 \simeq 3$ (for a 2D system) at a shock formation time $\tau_{\rm sf}\simeq 1030\,\omega_{\rm pi}^{-1}$(Fig.~\ref{Figure2}a), 
where $n_2$ is the compressed (downstream) density and $n_1$ is the single flow (upstream) density. We have verified that in 3D the density compression reaches $n_2/n_1 \simeq 4$ (see Appendix~\ref{app:3DandDiv}). In the homogeneous plasma case, the density compression observed is significantly weaker and slower. We have run a larger simulation with $L_x = 1200 \, c/\omega_{\rm pi}$ up to $2500\,\omega_{\rm pi}^{-1}$ and the maximum compression obtained was just $n_2/n_1 \simeq 2.5$ (Fig.~\ref{Figure2}a). 

The faster shock formation observed for laser-ablated inhomogenous plasma flows is due to the transition from small-angle scattering to magnetic reflection. In the case of homogeneous flows, the magnetic fields produced by the Weibel instability have a wavelength that is much smaller than the ion gyroradius of the flows, $\lambda_{B} \ll r_{i}$ (Fig.~\ref{Figure2}b). Ions slow down via many small-angle scatterings with the magnetic fields~\citep{Kato2008,Ruyer2016} and kink-like 
instabilities of the current filaments~\citep{Ruyer2018}, which operate on long time scales compared to the ion cyclotron frequency. 
A different dynamics is observed in the case of inhomogeneous flows 
where ions are efficiently confined near the mid-plane. This is a consequence of the temporal decrease of the flow velocity (and associated ion gyroradius) arriving at the interaction region, intrinsic to laser-ablated plasmas, $r_{i}(x=0, t) \propto v(x=0,t)\simeq L_0/(\tau_0+t)$ (
the $c_{\rm s} \ll v_0$ term 
is negligible in high-$M$ experiments). As the ion gyroradius becomes comparable to the dominant Weibel-mode wavelength, $r_{i} \sim \lambda_{B}$ (see Fig.~\ref{Figure2}b), the flows are effectively slowed down and heated by large-angle magnetic reflections.

\begin{figure}\centering
    \includegraphics[width=0.5\textwidth]{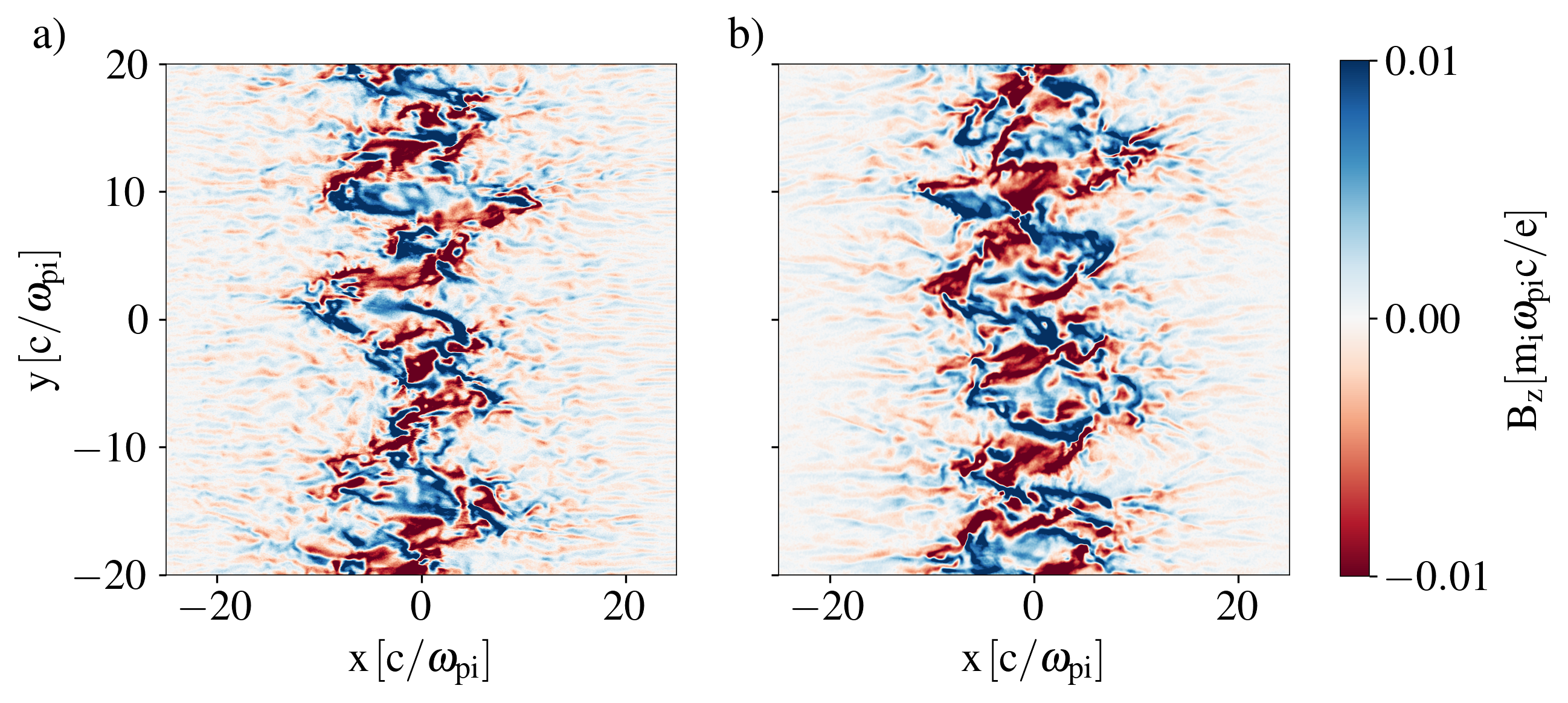}
     \caption{Magnetic field $B_z$ at $t\simeq 600\,\omega^{-1}_{\rm pi}$ for a) $v_{\rm th}= 0.02\, v_{\rm flow}$ and for b) $v_{\rm th}=1.25 \,v_{\rm flow}$. The electron temperature can be used to control the level of shock front corrugations.}
    \label{FigureS7}
\end{figure}

As the shock is formed in the inhomogeneous flow interaction, strong corrugations of the magnetic field (and density) are observed at the shock front on the scale of the ion gyroradius. 
The plasma flows primarily in alternating filaments near the mid-plane region. Weaker magnetic fields from the upstream region are advected towards the center through these filaments and slowed down on a scale comparable to the ion gyroradius. This leads to the compression of the magnetic fields near the mid-plane 
and to strong anti-symmetric modulations 
at the shock front, as observed in Fig.~\ref{Figure1}g. Since the magnetic field is frozen-in the electron fluid, the ratio of 
electron thermal 
to fluid velocity, $v_{th}/v_0$, controls the shock front corrugation level. 
Figure~\ref{FigureS7} shows that indeed the corrugations are significantly less pronounced when $v_{th} \gtrsim v_0$. 
Experimentally, $v_{th}$ 
can be controlled by varying the laser intensity and $Z$ 
(through radiative effects~\cite{Farley1999}) to probe 
the impact of shock front corrugations on 
magnetic field amplification and particle injection~\cite{Caprioli2014,Johlander2016,Marle2018}.

\section{Model of shock formation with laser-ablated flows}
A model for shock formation in laser-ablated counter-streaming plasma flows can be derived by estimating the critical time $\tau_{\rm c}$ at which the magnetic wavelength of the Weibel mode becomes comparable to the ion gyroradius at the interaction region, $r_{i}(\tau_{\rm c}) \sim \lambda_{B}(\tau_{\rm c})$. After saturation of the Weibel instability, the evolution of the dominant magnetic wavelength due to filament merging follows~\citep{Ruyer2015} $\lambda_{B}(t)\sim \lambda_{\rm sat}\left[1+((t-\tau_{\rm sat})/\tau_{\rm m})^2\right]$, where $\lambda_{\rm sat}\sim c/\omega_{\rm pi}$ is the wavelength at saturation, 
$\tau_{\rm m} \sim \frac{2\pi c}{v_0} \left(4m_i/Zm_e\right)^{1/4}\omega_{\rm pi}^{-1}$ is the typical merging time, and $v_0=L_0/\tau_0$ is the initial flow velocity. The ion gyroradius is $r_{i}(t) = m_i c v(t)/(Ze B_{\rm sat}) = 2\pi (c/\omega_{\rm pi}) L_0/(L_0 + v_0 t)$.
The critical time $\tau_{\rm c}$ 
is the solution of 
\begin{equation}\label{Equation3rdOrder}
   \left(1+ \frac{\tau_{\rm c}v_0}{L_0}\right) \left(  1+ \frac{(\tau_{\rm c}-\tau_{\rm sat})^2}{\tau_{\rm m}^2} \right)=2\pi \, .
\end{equation}
For the conditions of our simulation, this yields $\tau_{\rm c} \simeq 565\,\omega_{\rm pi}^{-1}$, in good agreement with our results (Fig.~\ref{Figure2}b). 

We 
note that filament merging ceases at late times for both homogeneous and inhomogeneous flows (Fig.~\ref{Figure2}b) due to the development of the drift-kink instability 
~\cite{Ruyer2018}. However, in practice, this  occurs at times comparable to or larger than $\tau_c$ and thus the use of the filament merging model to estimate $\tau_c$ is appropriate (see Appendix~\ref{app:Merging}). 

The critical time $\tau_{\rm c}$ marks the transition from small-angle scattering to magnetic reflection. At this stage, the ions reaching the center are slowed down and confined over a region of extension $\sim r_{i}$. The density compression in the mid-plane region between $\tau_{\rm c}$ and the shock formation time $\tau_{\rm sf}$ can then be described as 
\begin{equation}\label{equation1}
n_{\rm 2}(t) = 2n_{\rm 1}(\tau_{\rm c})+\int _{\tau_{\rm c}} ^{t} 2n_{\rm 1}(t')v(t')\frac{{\rm d}t'}{r_{i}(t')} \, ,
\end{equation}
where we have assumed that at $t=\tau_{\rm c}$ the two flows overlap without yet significant compression, i.e. $n_{\rm 2}(\tau_{\rm c})=2n_{\rm 1}(\tau_{\rm c})$. The shock formation time $\tau_{\rm sf}$ is 
obtained from the density jump condition $n_{\rm 2}(\tau_{\rm sf})=(2+\delta)n_{\rm 1}(\tau_{\rm sf})$, where $\delta=1$ in 2D and $\delta=2$ in 3D.

We have confirmed that Eq.~\ref{equation1} 
describes well the density compression observed in the simulations of laser-ablated plasma flows after $\tau_{\rm c}$, as shown in Fig.~\ref{Figure2}a. The predicted shock formation time $\tau_{\rm sf}\simeq 1060\,\omega_{\rm pi}^{-1}$ is also in good agreement with the simulation results. 
In the limit where the initial single flow density varies weakly over an ion gyroperiod, the shock formation time can be approximated by 
\begin{equation}\label{equation3}
\tau_{\rm sf} = \tau_{\rm c} +\frac{\delta}{2\Omega_c}\,
\end{equation}
with $\Omega_c = v(t)/r_{i}(t)$ the ion cyclotron frequency.

For typical experimental flow velocities 
~\cite{Kugland2012,Ross2012,Ross2013,Li2013,Fox2013,Huntington2015,Huntington2017,Ross2017} $v_0 \sim 1000-2000\, {\rm km/s} \ll c$, the critical time $\tau_{\rm c}$ largely exceeds the time of flow compression by magnetic reflection, and dominates the shock formation time $\tau_{\rm sf} \sim \tau_{\rm c}$. The minimum flow interpenetration distance required for shock formation is $L_{\rm sf} \sim \tau_{\rm sf} v_0 \sim \tau_{\rm c} v_0$ and thus corresponds to a minimum target separation $2 L_0 \sim 2 \tau_{\rm c} v_0$. By setting $\tau_{\rm c} = \tau_0$ and taking the appropriate limits $\tau_{\rm c} \gg \tau_{\rm m}$ and $\tau_{\rm c} \gg \tau_{\rm sat}$ we can 
calculate the minimum shock formation time as $\tau_{\rm sf} \simeq \tau_{\rm c} \simeq 16 (c/v_0)\left[m_i/(Z m_e)\right]^{1/4}\omega_{\rm pi}^{-1} \simeq 2.5 \left[m_i/(Z m_e)\right]^{1/4}\Omega_{\rm c}^{-1}$. 
In practice, laser-ablated fully ionized plasmas have $A/Z \simeq 2$, for which the required target separation to reach shock formation is $2 L_0 \simeq 250\, c/\omega_{\rm pi}$ and is independent of the flow velocity $v_0$.

\begin{figure}
    \centering
    \includegraphics[width=0.4\textwidth]{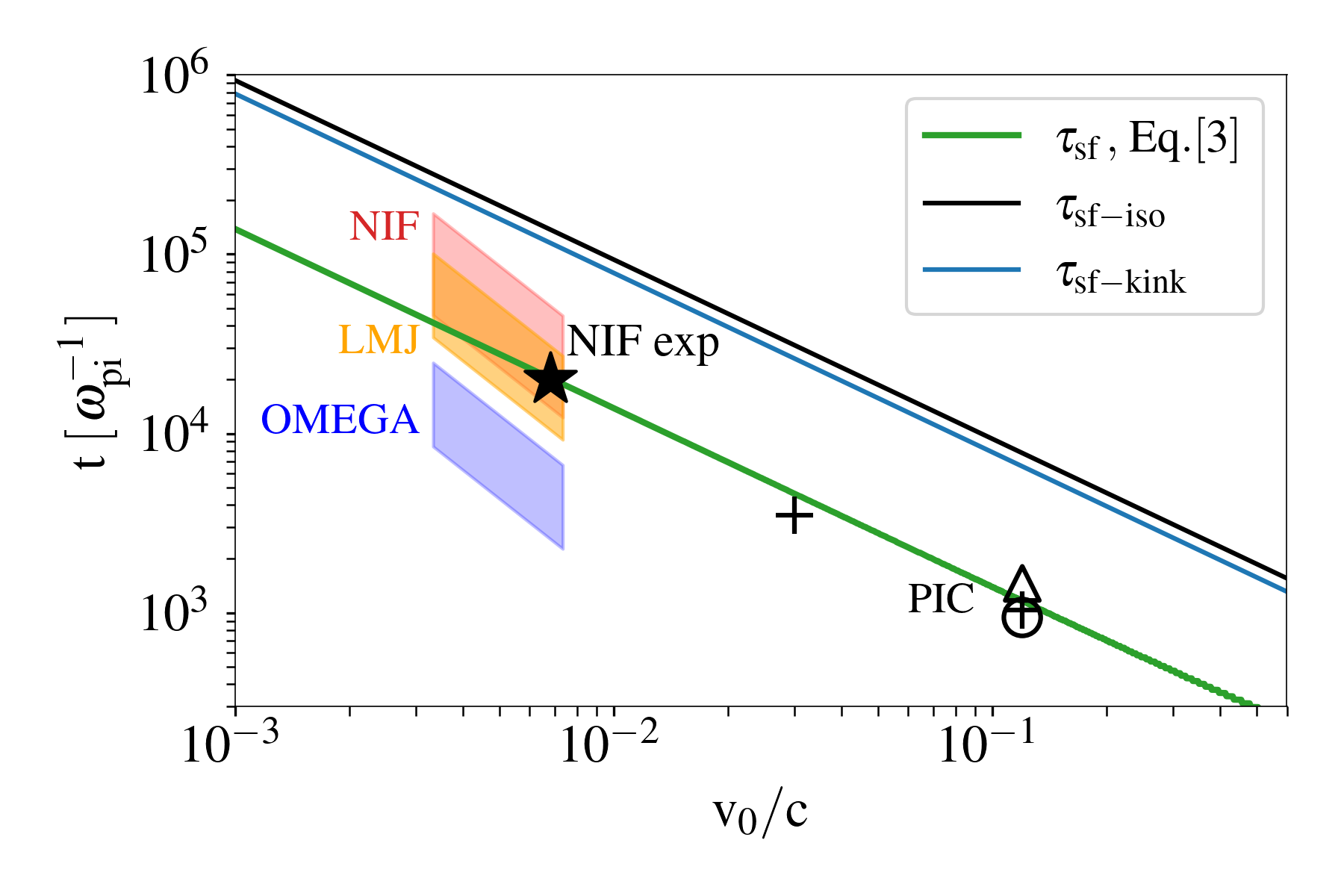}
    \caption{Comparison of the shock formation time $\tau_{\rm sf}$ from Eq.~\ref{equation3} (green line) for laser-ablated flows with $m_i/(Z m_e) = 3672$ and $L_0 = 125\,c/\omega_{\rm pi}$ with the results of 2D PIC simulations with  $m_i/(Z m_e) = 128$ ($ + $) and $m_i/(Z m_e) = 512$ ($ \triangle $), 3D simulations with $m_i/(Z m_e) = 32$ ($ \circ $) and with the experimental observation from Ref.~\citep{Fiuza2020} ($ \bigstar $).
    The shock formation times predicted for homogeneous flows 
    $\tau_{\rm sf-iso}$ (black line)~\citep{Ruyer2016} and 
    $\tau_{\rm sf-kink}$ (blue line)~\citep{Ruyer2018} largely exceed that observed for laser-ablated 
    flows. 
    The shaded areas indicate the range of parameters achievable at current HED laser facilities.}
    \label{Figure3}
\end{figure}
This analysis considered the longitudinal inhomogeneity of laser-ablated planar plasma flows. On time scales longer than $L_0/c_s$, 3D divergence effects can modify the plasma density profiles. It is thus important to guarantee that the shock formation time $\tau_\mathrm{sf} = L_\mathrm{sf}/v_0 < L_0/c_s $ or, equivalently, $L_0 > L_\mathrm{sf}/M$. This condition is always comfortably met for system sizes large enough to study high-Mach number shocks ($L_0 \ge L_\mathrm{sf}$ and $M \gg 1$) and thus the profiles considered are reasonably justified.
The transverse inhomogeneity of the plasma and 
obliquity of the flows away from their axes (due to divergence) can also affect 
plasma interpenetration. 
These effects 
should not impact the physics of shock formation and 
corrugation (both on the ion gyroradius scale) 
if $L_0 \gg r_i \sim 0.064/\sqrt{n_0 [10^{19} \mathrm{cm}^{-3}]}$ cm. 
For the typical system size, $L_0 \sim 0.4 - 1.25$ cm, and 
plasma density, $n_0 \sim 5\times10^{18} - 5\times10^{19} \mathrm{cm}^{-3}$, of current experiments~\cite{Ross2012,Huntington2017,Ross2013,Fiuza2020}, this condition is always met. 
We have performed additional 2D simulations with spherically divergent flows and 3D simulations with modified density profiles and confirmed that 
the shock formation time and structure 
predicted by our model are robustly observed in all cases (see Appendix~\ref{app:3DandDiv}).

\section{Comparison of shock formation models with PIC simulations and experiments}
Shock formation in counter-streaming laser-ablated plasmas is significantly faster than that predicted from previous models of Weibel-mediated shocks in homogeneous flows. The predicted shock formation time due to flow isotropization during filament merging is~\citep{Ruyer2016} $\tau_{\rm sf-iso} \simeq 35(m_i/Zm_e)^{0.4} (c/v_0) \omega_{\rm pi}^{-1}$, corresponding to 
$2 L_0 \simeq 1867 c/\omega_{\rm pi}$. Additionally, shock formation due to the disruption of the current filaments by kink-type instabilities 
predicts~\citep{Ruyer2018} $\tau_{\rm sf-kink} \simeq 5 \tau_{\rm kink} \simeq (5/3)(m_i/Zm_e)^{3/4} (c/v_0) \omega_{\rm pi}^{-1}$, where $\tau_{\rm kink}$ is the growth time of the drift-kink instability, and corresponds to 
$2 L_0 \simeq 1572 c/\omega_{\rm pi}$. The shock formation model proposed in this work is thus $6-8\times$ faster than previous predictions.

Figure \ref{Figure3} summarizes the comparison of our model predictions for the shock formation time and those of previous models~\cite{Ruyer2016,Ruyer2018} with the results of 2D and 3D PIC simulations and experimental measurements on NIF~\cite{Fiuza2020}. The PIC simulations performed over a range of flow velocities ($v_0 = 0.03 - 0.1 \,c$) and mass to charge ratios $[m_i/(m_e Z) = 32 - 512]$ show good agreement with our model, confirming the scaling of $\tau_{\rm sf} \propto 1/v_0$ (shock formation distance 
independent of $v_0$) and the weak dependence on mass to charge ratio. Moreover, our model is in good agreement with the recent NIF experiments, 
which observed shock formation
after $\lesssim 3$ ns of interaction. For the measured plasma overlapping density 
$\simeq 5\times 10^{19}\rm cm^{-3}$ and 
$v_0 \simeq 2000\,\rm km/s$, the shock formation time predicted by our model is $2\times10^4\,\omega_{\rm pi}^{-1}\simeq 3\,\rm ns$, consistent with the experimental measurements.

The results presented here are important to guide the development and interpretation of experimental studies of turbulent collisionless shocks and in connecting them with astrophysical simulations and models. Figure ~\ref{Figure3} indicates the 
parameter space (shaded areas) that can be probed by the current HED facilities (see Appendix~\ref{app:HEDexp} for details). For kJ-class OMEGA experiments~\cite{Kugland2012,Ross2012,Ross2013,Li2013,Fox2013,Huntington2015,Huntington2017}, with measured plasma densities $n_0 \sim 10^{18} -  10^{19} {\rm cm}^{-3}$, 
we predict $\tau_{\rm sf} \gtrsim 13$ ns and 
$2 L_0 \gtrsim 2.5$ cm, which largely exceed the interaction time of $\sim 5$ ns and system size of $\sim 0.8$ cm of previous 
experiments. 
However, our results indicate that shock formation can be comfortably reached at higher energy laser facilities ($> 100$ kJ), such as NIF and LMJ. 
The collisionless shocks will have 
a transition size dictated by the ion gyroradius and the magnetic turbulence in the shock foot is driven by the Weibel instability, which are critical properties of high-Mach number astrophysical shocks revealed by kinetic simulations~\cite{Matsumoto2017} and spacecraft observations~\cite{Sundberg2017}. Moreover, by varying the electron temperature it will be possible to control 
and study the impact of shock front 
corrugations on magnetic field amplification and particle injection~\cite{Caprioli2014,Johlander2016,Marle2018}. 

\section{Conclusions}
In summary, we have shown for the first time that the inhomogeneity of laser-ablated plasma flow profiles has a significant impact in accelerating the formation of turbulent collisionless shocks and in controlling their dynamics. These findings have important implications as they enable the use of current HED experimental facilities to study the complex interplay between magnetic field amplification, shock front corrugation, and particle injection underlying high-Mach number astrophysical shocks.

The authors thank D. D. Ryutov, D. P. Higginson, and H.-S. Park for stimulating discussions. This work was supported by the U.S. Department of Energy SLAC Contract No. DEAC02-76SF00515 and by the U.S. DOE Early Career Research Program under FWP 100331. The authors acknowledge the OSIRIS Consortium, consisting of UCLA and IST (Portugal) for the use of the OSIRIS 3.0 framework and the visXD framework. Simulations were run on Mira (ALCF) through an ALCC award and on Cori (NERSC).

\appendix
\section{Initialization of the laser-ablated plasma profiles}\label{app:Init}
\begin{figure}
    \includegraphics[width=0.4\textwidth]{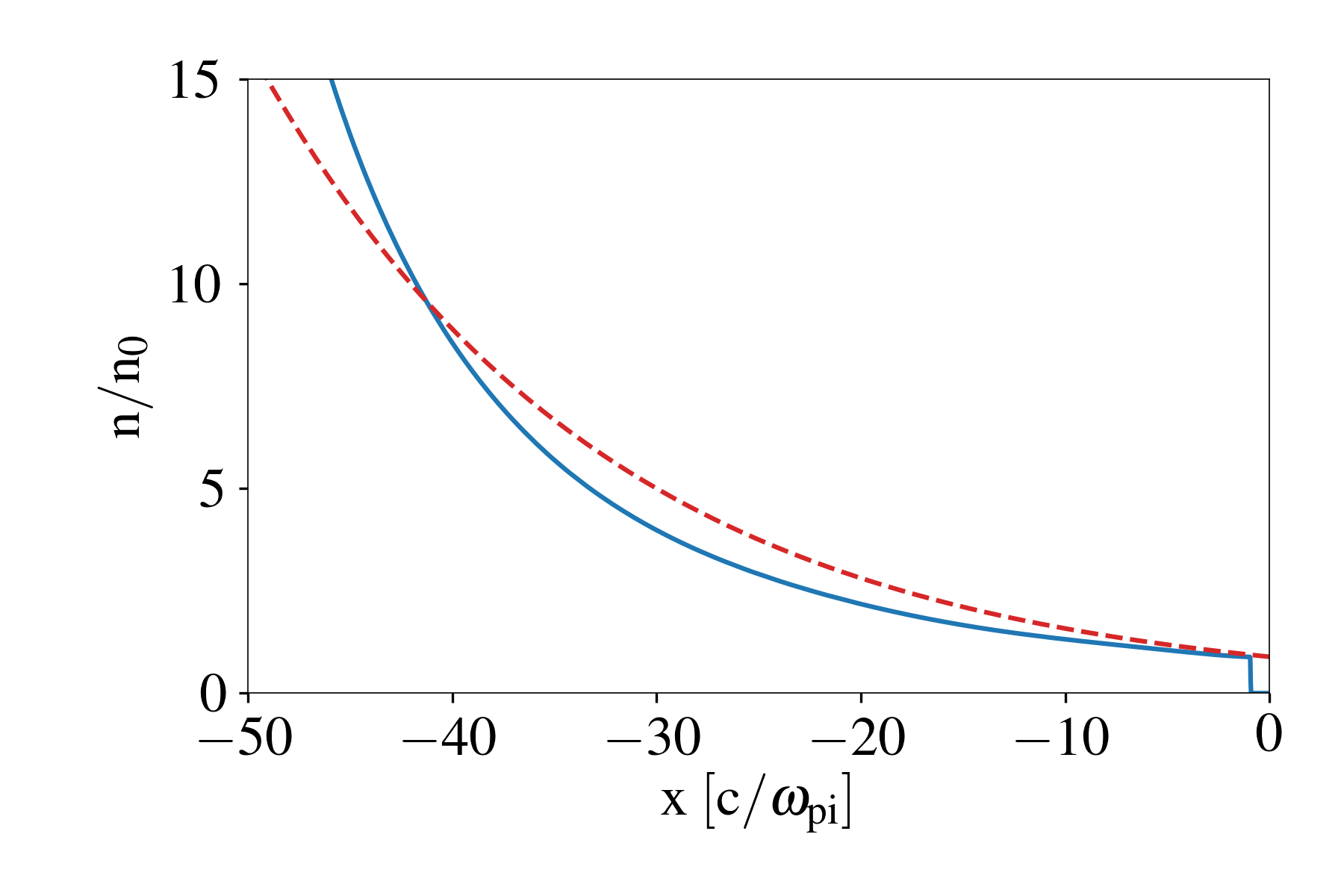}
    \caption{Comparison of the density profile of the forward moving inhomogeneous flow initialized in the simulations (blue curve) and that predicted from self-similar theory ~\citep{Gurevich1966} (red dashed curve).}
    \label{FigureS1}
\end{figure}

The plasma used in our inhomogeneous simulations has a velocity profile $v(x,t)= c_{\rm s} + (x+L_0)/(t+\tau_0)$ that follows the self-similar solution for laser-ablated plasmas \citep{Gurevich1966}. The density profile has been slightly modified from the self-similar solution in order to guarantee that the flow energy density, $n v^2$, delivered at the mid-plane region (where the flows interact) is constant in time $n v^2=n_0 v_0^2$. This is convenient because it allows for a direct comparison of the shock formation efficiency between inhomogeneous (laser-ablated) and homogeneous flows. We compute the initial density profile $n(x,t=0)$ by integrating the continuity equation for the density, with boundary condition $n(x=0,t)$ and assuming ballistic propagation of a density element (i.e. ${\rm d}v/{\rm d}t=0$). This gives the density profile showed in Fig.~\ref{FigureS1} (blue line), which is a reasonable approximation of the self-similar profile (red dashed line). We have performed simulations with different density profiles and checked that the main results of this work are relatively insensitive to the exact details of the density profile and determined primarily by the inhomogeneity of the flow velocity, which is responsible for the transition from small-angle scattering to magnetic reflection in the shock formation.

\section{Growth rate and saturation of the Weibel instability in inhomogeneous counter-streaming flows}\label{app:WeibelGrowth}
The growth of the ion Weibel instability in inhomogeneous counter-streaming flows can be best described in a frame where the instability is purely growing (termed "Weibel frame"), defined as
\begin{equation}\label{WeibelFrame}
u_{\rm wf } = \frac{\frac{n_+v_+}{T_+} + \frac{n_-v_-}{T_-}}{\frac{n_+}{T_+} + \frac{n_-}{T_-}}\, ,\tag{$\rm S1$}
\end{equation}
\begin{figure}
    \includegraphics[width=0.4\textwidth]{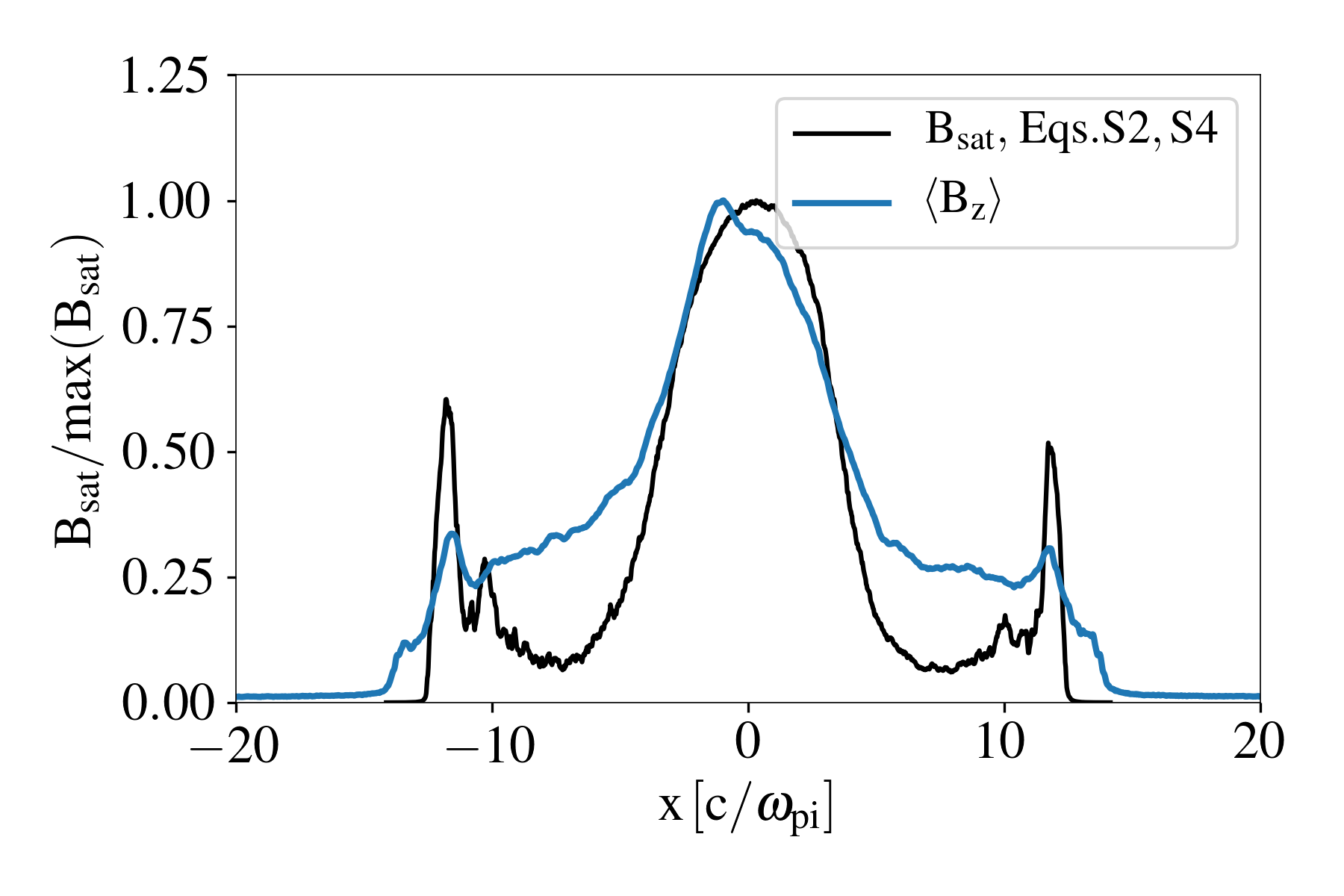}
    \caption{Comparison of the longitudinal spatial profile of the magnetic field measured in the inhomogenous flow simulation at $t=125\,\omega_{\rm pi}^{-1}$ (blue curve) with the prediction of the magnetic trapping mechanism (Eq.~\ref{Bsat}) for the growth rate obtained from Eq.~\ref{DispertionRelation} (black curve).}
    \label{FigureS2}
\end{figure}

with $n_{\pm},v_{\pm},T_{\pm}$ the local densities, velocities, and temperatures of the ion flows moving in the positive (+) and negative (-) direction. The corresponding dispersion relation, accounting for the ion dynamics in the non-relativistic regime, reads
\begin{equation}\label{DispertionRelation}
\omega^2 -c^2k^2 + {\omega_{\rm pi}^*}^2 -k^2\left[ \frac{\omega_{+}^2v_+'^2}{\omega^2-k^2v_{\rm th,+}^2}+\frac{\omega_{-}^2v_-'^2}{\omega^2-k^2v_{\rm th,-}^2} \right] \,=0 ,\tag{$\rm S2$}
\end{equation}
where $\omega_{\rm pi}^*$ is the ion plasma frequency at the local density, $v_{\rm th}$ is the ion thermal velocity and the primed quantities are defined in the Weibel frame. The growth rate $\Gamma=i\omega$, obtained from the solution of Eq.~\ref{DispertionRelation}, in the limit of cold flows and $k\gg \omega_{\rm pi}^*/c$, can be expressed as 
\begin{equation}\label{WeibelGrowthLAB}
\Gamma = \frac{\sqrt{n_+n_-}}{n_++n_-}\left(v_+-v_-\right)\omega_{\rm pi}^* \, ,\tag{$\rm S3$}
\end{equation}
which is maximum at the center of interaction ($x=0$), where the flows have equal density and opposite velocity, consistent with the simulation results. However, it is important to take into account temperature effects, since the heating of the flows as they cross the strong field region near the mid-plane will further decrease the growth rate away from it. This will also impact the saturation of the magnetic field, which, following the magnetic trapping mechanism ~\citep{Davidson1972}, is given by 
\begin{equation}\label{Bsat}
B_{\rm sat}= \frac{\Gamma^2}{2\pi v_0} \frac{m_i c^2}{e \omega_{\rm pi}} \, . \tag{$\rm S4$}
\end{equation}
We have compared the magnetic field profile obtained in the inhomogeneous simulation with that predicted by Eqs.~\ref{DispertionRelation} and~\ref{Bsat}, where we have solved Eq.~\ref{DispertionRelation} numerically for the ion temperature, density, and velocity extracted from the simulation. The results, shown in Fig.~\ref{FigureS2} near the saturation of the instability ($t=125\,\omega_{\rm pi}^{-1}$), indicate that the sharp decrease of the magnetic field amplitude away from the mid-plane in inhomogeneous flows can be reasonably well described by our model.

\section{Interplay between filament merging and kink instability}~\label{app:Merging}
\begin{figure}
    \includegraphics[width=0.4\textwidth]{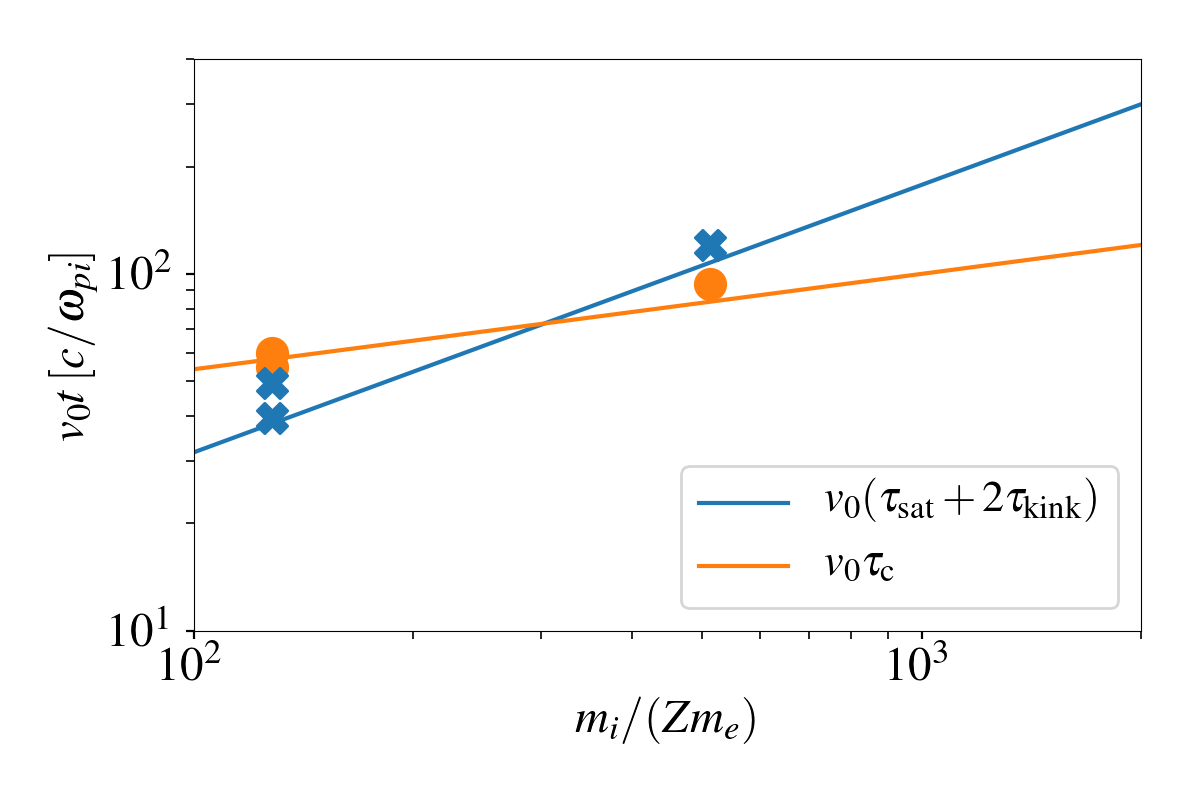}
    \caption{Comparison of the time at which the filament wavelength matches the ion gyroradius (orange dots) and the filament merging saturates (blue crosses) observed in PIC simulations with the estimates of $\tau_c$ (Eq. 1) and $\tau_{\rm kink}$ \cite{Ruyer2018}. The PIC simulations with $m_i/(Z m_e) = 128$ have $v_0 = 0.11\,\rm c$ and $v_0 = 0.03\,\rm c$, and the simulation with $m_i/(Z m_e) = 512$ has $v_0 = 0.11\,\rm c$.}
    \label{FigureS3}
\end{figure}

Our PIC simulations (for both homogeneous and inhomogeneous flow profiles) show that right after the saturation of the Weibel instability the dynamics of the current filaments is primarily governed by filament merging, in good agreement with the model of Ref. \citep{Ruyer2016} (see Fig. 2b). However, at late times ($t\simeq 500\,\omega_{\rm pi}^{-1}$ in Fig. 2b), we observe that filament merging ceases and the dominant magnetic wavelength remains approximately constant. We have found that the stopping of the filament merging coincides with the development of longitudinal modulations of the filaments and the onset of magnetic turbulence, consistent with the development of kink-like instabilities described in Ref.~\citep{Ruyer2018}. We observe that this transition occurs at a time $t \simeq 2\tau_{\rm kink}\simeq (m_i/Zm_e)^{3/4}(c/v_0)\,\omega_{\rm pi}^{-1}$ after the saturation of the Weibel instability ($\tau_{\rm sat}$), which is consistent across the range of simulations performed for different flow velocities and ion to electron mass ratios (Fig.~\ref{FigureS3}). We have also verified that our estimate of the critical time $\tau_c$ for the transition between small-angle scattering and magnetic reflection (Eq. 1) is in good agreement with the PIC simulation results (Fig.~\ref{FigureS3}). We observe that for realistic ion to electron mass ratios the critical time occurs always before the saturation of filament merging (onset of the kink instability), thus validating the use of the merging model in the estimate of $\tau_c$.

\section{Impact of flow divergence and 3D effects on shock formation}\label{app:3DandDiv}
Laser-produced plasma flows will have an associated divergence, \emph{i.e.} they will have a radial velocity component away from the central axis ($r = 0$). The flow divergence at the interaction region can be controlled by the ratio of the laser focal spot size and the interaction distance $L_0$. Earlier experiments used laser spot sizes significantly smaller than the system size, and the flow divergence was relatively large ~\cite{Ross2012,Ross2013,Huntington2015,Ross2017}. Very recently, experiments at the National Ignition Facility (NIF) ~\cite{Fiuza2020} used a larger spot size comparable to the system size, and showed that the transverse size of the heated plasma region at the flow-flow interaction is comparable to the laser spot at the target, indicating that the flows are relatively well collimated, as considered in our simulations. 
In general, as discussed in the main text, provided that the system size is much larger that the ion gyroradius associated with the Weibel magnetic field, $L_0\gg r_i$, the flow divergence is not expected to affect the shock formation physics and the use of collimated flows considered in the simulations is well justified.

\begin{figure}
    \includegraphics[width=.5\textwidth]{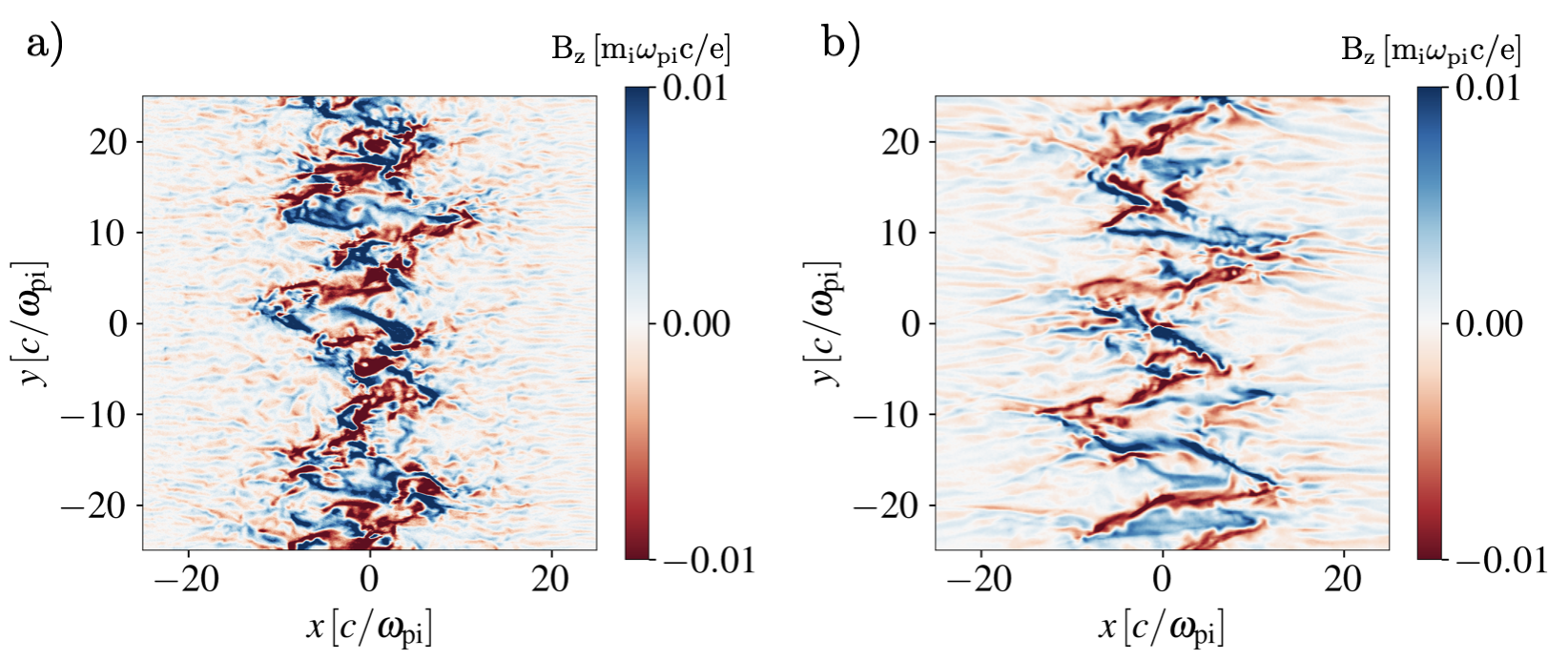}
    \caption{Magnetic field $B_z$ at $t\simeq 570\,\omega^{-1}_{\rm pi}$
for a) the simulation with collimated flows as in Fig. 1d and b) a case with spherical flow expansion.}
    \label{FigureS8}
\end{figure}

In order to explore the impact of the plasma flow divergence, we performed a simulation with spherically divergent flows. The density and velocity profiles along the $y=0$ axis are equal to the ones discussed in the manuscript and constant at a radial distance $R$ from the target irradiated spot. The flow is transversely limited at $\pm 30^{\circ}$ from the $y=0$ axis 
to minimize the need of a very large transverse simulation box size.
As can be seen in Fig.~\ref{FigureS8}, the simulation with spherically expanding flows (right panel) shows a magnetic field amplitude and development of corrugations similar to the one presented in the main paper (left panel). This confirms that the dynamics is weakly affected by the
flow divergence if the ion gyroradius is much smaller than the system size, as in the recent NIF experiments.

\begin{figure}
    \includegraphics[width=.4\textwidth]{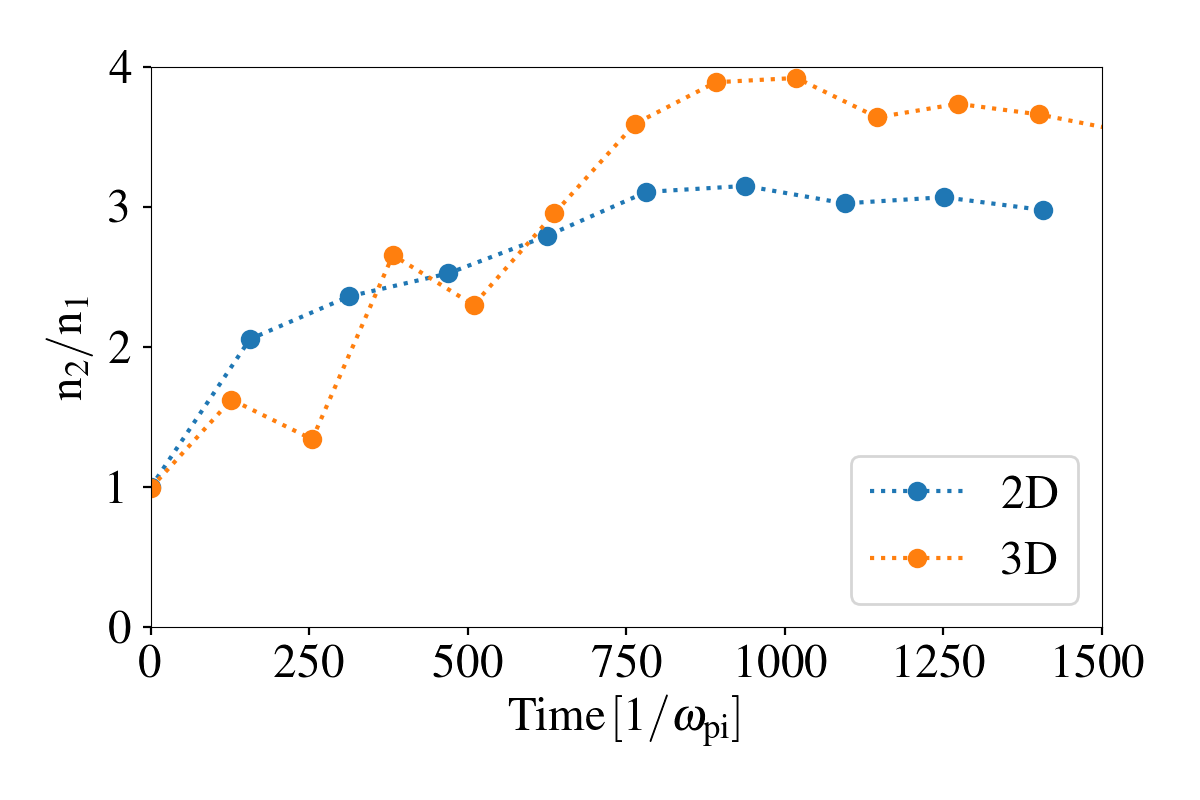}
    \caption{Evolution in time of the compression factor at the shock front for the 2D simulation (blue dots) and the corresponding 3D simulation (orange dots) with $m_i/(Zm_e)=32$. }
    \label{FigureS6}
\end{figure}

In addition, we have performed 3D simulations to verify that the shock physics and model prediction for the shock formation time is robust in 3D. We have used $m_i/(Zm_e)=32$ due to the very high computational cost. The velocity and density profiles are the same as in Fig. 1, with the difference that the density is kept constant after $\tau_{\rm plateau}=700\,\omega_{\rm pi}^{-1}$.
This plateau in density profile is typically observed experimentally due to the flow divergence and/or finiteness of the laser duration \cite{Ross2012,Fiuza2020}. In order to confirm that this does not impact significantly the shock formation physics, which is primarily determined by the flow velocity profile, we have simulated a case where $\tau_{\rm sf}>\tau_{\rm plateau}$. As can be seen in Fig.~\ref{FigureS6}, shock formation is reached in 3D between $t \sim 900 - 1000 \,\omega_{\rm pi}^{-1}$ with a density compression factor of 4, as expected from the shock jump conditions. The observed shock formation time is in very good agreement with our model prediction of $\tau_{\rm sf}\simeq 940\,\omega_{\rm pi}^{-1}$.

\section{Plasma conditions of current HED facilities}~\label{app:HEDexp}
There has been a significant experimental effort in using high-energy-density laser facilities such as OMEGA and NIF to explore the physics of collisionless shocks in counter-streaming laser-ablated plasmas \citep{Kugland2012,Ross2012,Ross2013,Li2013,Fox2013,Huntington2015,Huntington2017,Ross2017,Swadling2020,Fiuza2020}. These experiments produce typical flow velocity $v_0\sim 1000-2000$ km/s, with the plasma density being primarily constrained by the laser energy and system size. Experiments at OMEGA~\citep{Kugland2012,Ross2012,Ross2013,Li2013,Fox2013,Huntington2015,Huntington2017} have been performed with up to 5kJ/target and have measured typical densities at the interaction region in the range $10^{18}-10^{19} \,\rm cm^{-3}$. Recent experiments on the NIF used 250-450 kJ/target ~\citep{Ross2017,Fiuza2020} and measured a density at the interaction in the range $10^{19} - 10^{20}\,\rm cm^{-3}$. While, to our knowledge, collisionless shock experiments on LMJ have not yet been reported, based on the results from NIF, at the currently available energy of 150 kJ/target it is reasonable to expect plasma densities in the range $5\times 10^{18} - 5\times 10^{19}\,\rm cm^{-3}$. The maximum plasma system sizes that can be produced at these conditions can be estimated from basic energy conservation arguments as
\begin{equation}\label{system_size}
\epsilon_{\rm laser} = \frac{\eta}{2} n_0 m_i v_0^2 \frac{L_0^3}{\tan(\theta)^2}\, ,\tag{$\rm S5$}
\end{equation}
where $\eta$ is the energy conversion efficiency from the laser to the plasma flows and $\theta$ is the divergence angle of the flows. We have considered typical parameter ranges $\eta \sim 0.25-0.5$ and $\theta \sim 30^\circ - 45^\circ$. The range of interaction times $L_0/v_0$ obtained is shown in Fig. 3.

To study collisionless shocks, it is important to guarantee that Coulomb collisions will not affect the shock formation process. For collisions to have a negligible effect, one should guarantee that the ion-ion mean free path 
$\lambda_{\rm mfp} = m_i^2 v_r^4/(16 \pi Z^4 e^4 n_i \log\Lambda)$, which is responsible for slowing down the flows, is much larger than the required shock formation distance $L_{\rm sf} = 125\, c/\omega_{\rm pi}$ (as described in our model). Here $n_i = n_0/Z$ is the ion density of each flow, $v_r = 2 v_0$ is the relative velocity between flows, and $\log\Lambda$ is the Coulomb logarithm. This condition can be written as 
\begin{equation}
\frac{\lambda_{\rm mfp}}{L_{\rm sf}} \sim  \frac{74}{Z}\frac{(v_0 {[\rm 1000\, km/s]})^4}{\sqrt{n_0 [\rm 10^{19}\, cm^{-3}]}} \gg 1, \tag{$\rm S6$}
\end{equation}
where we have used $A/Z=2$ and $\log\Lambda=10$, as typical of laser-ablated plasmas. 
We find that for the typical range of flow density and velocity values of experiments discussed above and $Z = 6$ (Carbon flows), we have $\lambda_{\rm mfp}/L_{\rm sf} \sim 6 - 280$ and thus collisions are not expected to affect the shock formation process discussed in our work.

\bibliographystyle{unsrt}
\bibliography{Biblio}

\end{document}